\documentclass[sigconf]{acmart} 
\usepackage{color}
\usepackage{amsmath}
\usepackage{amsthm}
\usepackage{amsfonts}
\usepackage{algorithm}
\usepackage{algorithmic}
\usepackage{graphicx}
\usepackage{multirow}
\usepackage{subfigure}
\usepackage[normalem]{ulem}
\usepackage{booktabs}

\useunder{\uline}{\ul}{}

\usepackage{array}
\newcolumntype{L}[1]{>{\raggedright\let\newline\\\arraybackslash\hspace{0pt}}m{#1}}
\newcolumntype{C}[1]{>{\centering\let\newline  \\\arraybackslash\hspace{0pt}}m{#1}}
\newcolumntype{R}[1]{>{\raggedleft\let\newline \\\arraybackslash\hspace{0pt}}m{#1}}
\graphicspath{ {img/} }

\def\bH{\mathbf{H}}

\def\mM{\mathcal{M}}

\def\bC{\mathbf{C}}

\def\bW{\mathbf{W}}

\def\cA{\mathcal{A}}

\def\cP{\mathcal{P}}

\def\cV{\mathcal{V}}

\theoremstyle{definition}

\begin{document}

\title{Motif Enhanced Recommendation over Heterogeneous Information Network}

\author{Huan Zhao}
\authornote{Huan Zhao is currently affiliated with 4Paradigm Inc.}
\author{Yingqi Zhou, Yangqiu Song, Dik Lun Lee}
\affiliation{	
	\institution{Department of Computer Science and Engineering}
	\institution{Hong Kong University of Science and Technology}
	\city{Kowloon}
	\country{Hong Kong}
}
\email{hzhaoaf@cse.ust.hk;yzhoubb@connect.ust.hk;{yqsong,dlee}@cse.ust.hk}


\copyrightyear{2019} 
\acmYear{2019} 
\acmConference[CIKM '19]{The 28th ACM International Conference on Information and Knowledge Management}{November 3--7, 2019}{Beijing, China}
\acmBooktitle{The 28th ACM International Conference on Information and Knowledge Management (CIKM '19), November 3--7, 2019, Beijing, China}
\acmPrice{15.00}
\acmDOI{10.1145/3357384.3358134}
\acmISBN{978-1-4503-6976-3/19/11}



\renewcommand{\shortauthors}{Huan Zhao et al.}

\begin{abstract}
Heterogeneous Information Networks (HIN) has been widely used in recommender systems (RSs). In previous HIN-based RSs, meta-path is used to compute the similarity between users and items. However, existing meta-path based methods only consider first-order relations, ignoring higher-order relations among the nodes of \textit{same} type, captured by \textit{motifs}. In this paper, we propose to use motifs to capture higher-order relations among nodes of same type in a HIN and develop the motif-enhanced meta-path (MEMP) to combine motif-based higher-order relations with edge-based first-order relations. With  MEMP-based similarities between users and items, we design a recommending model MoHINRec, and experimental results on two real-world datasets, Epinions and CiaoDVD, demonstrate its superiority over existing HIN-based RS methods.

\end{abstract}

\maketitle

\section{Introduction}

\begin{figure}
	\centering
	\subfigure[Example HIN.]{\includegraphics[width=0.22\textwidth]{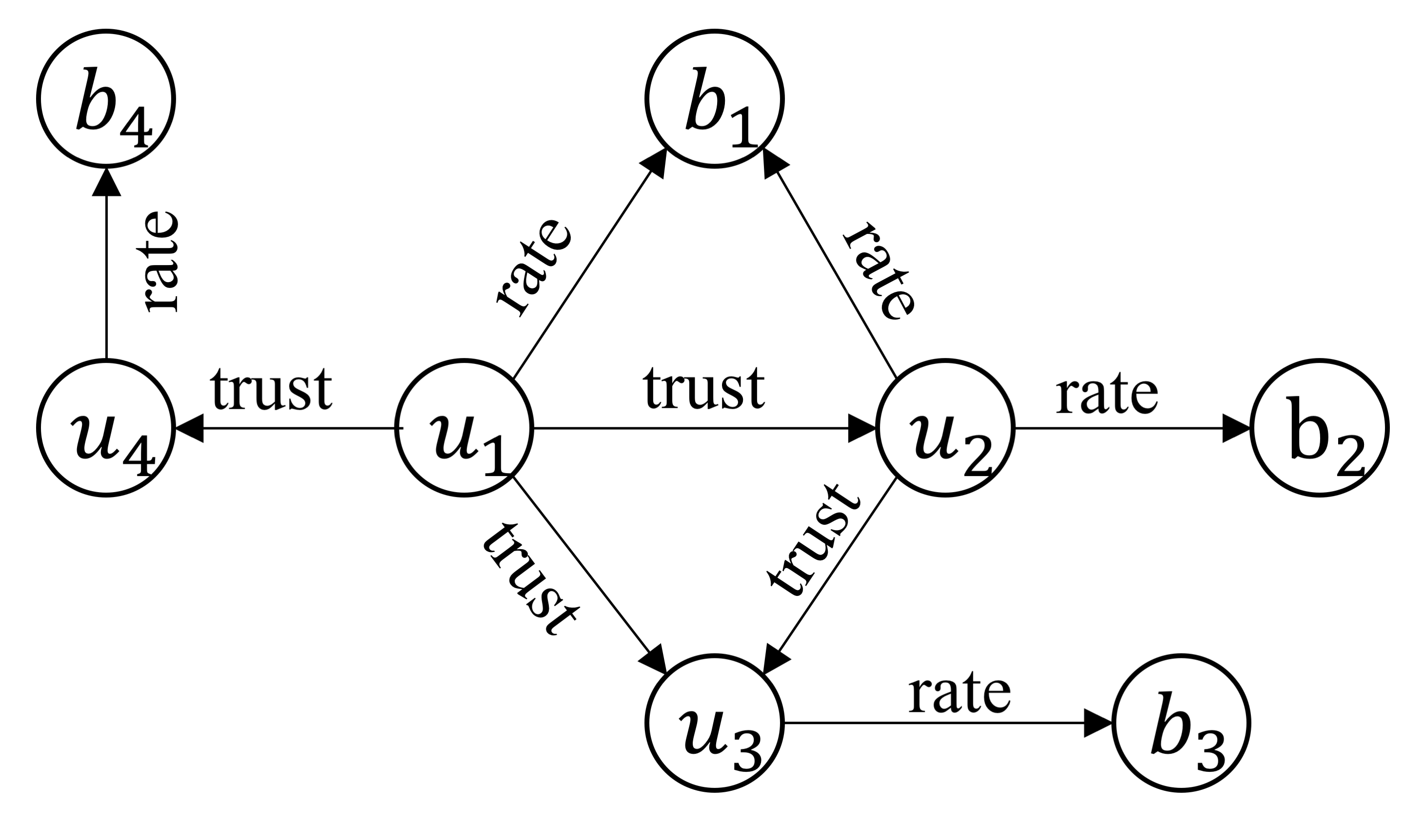}}\hspace{0.1in}
	\subfigure[Meta-paths.]{\includegraphics[width=0.22\textwidth]{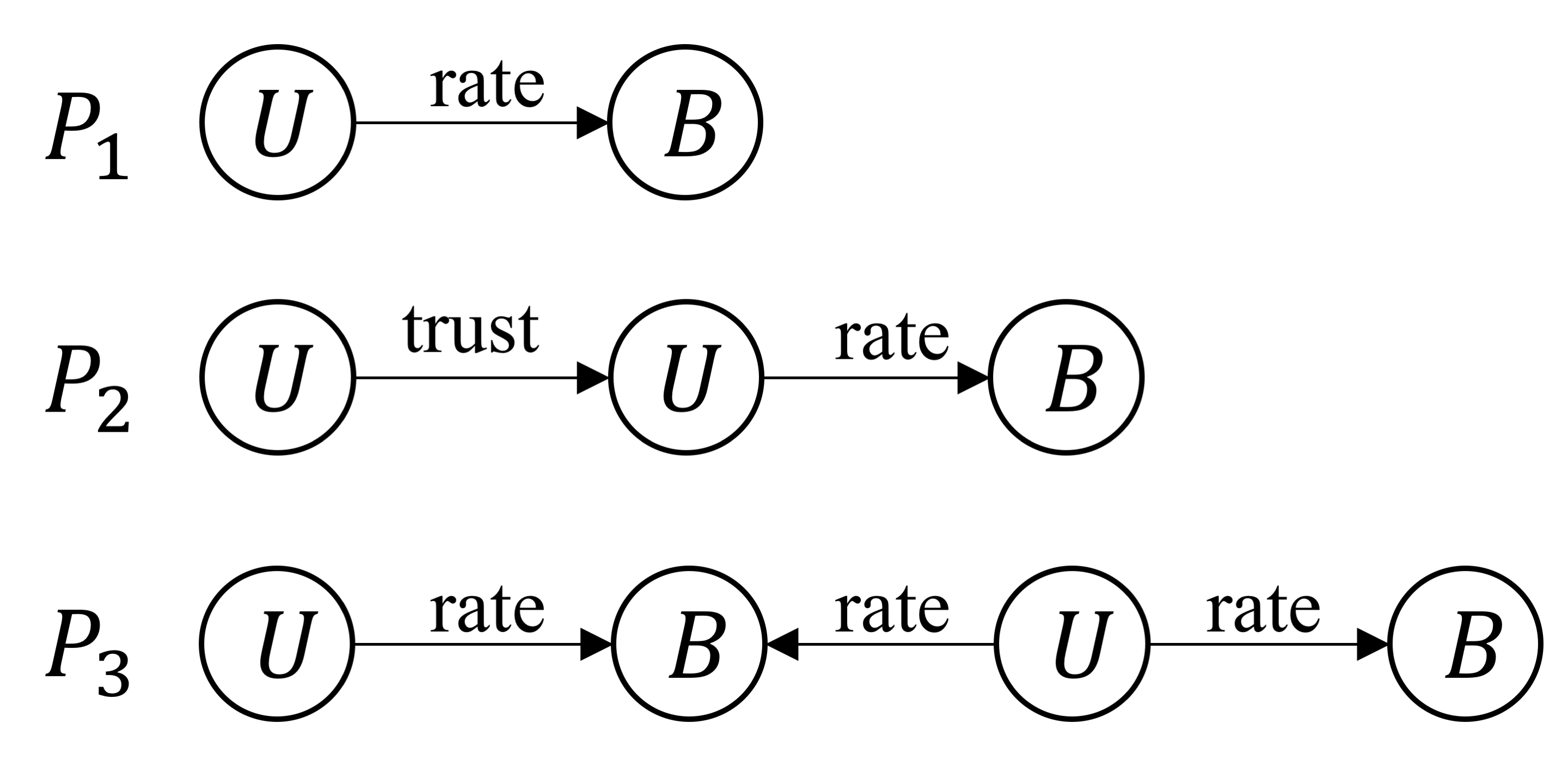}}
	\caption{An example HIN extracted from the Epinions dataset. We also show the network schema and the corresponding meta-paths. Note that we omitted other SIs such as category and review texts in (a).}
	\label{fig-example-hin}
	\vspace{-0.2in}
\end{figure}

Heterogeneous Information Network (HIN)~\cite{sun2011pathsim} has been a popular framework in Recommender Systems (RSs) for its capability to model all sorts of heterogeneous side informations (SIs), which can improve recommending performance~\cite{yu2014personalized,shi2015semantic,zhao2017meta,shi2018heterogeneous,han2018aspect,Hu2018LMB,Wang2018Billion,Fan2019MHG}. 
Most of the existing HIN-based RS methods utilize the number of instances of meta-path, a sequence of node types in a HIN, connecting the users and items in computing similarity: the larger the number, the higher the similarity. For example, for $\cP_2$ in Figure~\ref{fig-example-hin}(b), the number of meta-path instances connecting $u_1$ and $b_2$ in Figure~\ref{fig-example-hin}(a) is $1$, i.e., $u_1 \rightarrow u_2 \rightarrow b_2$ is the only meta-path instance of $\cP_2$ connecting $u_1$ and $b_2$. In~\cite{sun2011pathsim}, commuting matrix was proposed to represent the number of meta-path instances connecting two nodes, which can be obtained by multiplying a series of adjacency matrices built on pairs of node types. For example, the commuting matrix $\bC_{\cP_2}$ for meta-path $\cP_2$ can be obtained by $\bC_{\cP_2} = \bW_{UU} \cdot \bW_{UB}$. Each entry in $\bC_{\cP_2}$ represents the number of $\cP_2$ instances connecting the users and the items, and $\bW_{UU}$ and $\bW_{UB}$ are the adjacency matrices for the corresponding pairs of types, i.e., (\textit{User}, \textit{User}) and (\textit{User}, \textit{Item}), respectively. 
The meta-path based similarities are then used as features in different recommending models.

In this paper, we argue that existing HIN-based RS methods have not fully exploited all of the information available in HIN. Consider Figure~\ref{fig-example-hin}(a), the similarities for the pairs $(u_1, b_3)$ and $(u_1, b_4)$ based on $\cP_2$ in Figure~\ref{fig-example-hin}(b) are both 1, because both pairs have only one $\cP_2$ instance connecting their end nodes (respectively, $u_1 \rightarrow u_3 \rightarrow b_2$ and $u_1 \rightarrow u_4 \rightarrow b_4$). Technically speaking, $\cP_2$ models social recommendation, i.e., if we want to recommend items to $u_1$ based on $u_1$'s friends, $b_3$ and $b_4$ will be the same because they have the same similarity to $u_1$. However, when examining Figure~\ref{fig-example-hin}(a) carefully, we can see that it is better to recommend $b_3$ over $b_4$ to $u_1$ because $u_1$ trusts $u_3$ more than $u_4$. This is because $u_1$ trusts $u_3$ not only directly by its link to $u_3$ but also indirectly via $u_2$. In other words, $u_1$, $u_2$, and $u_3$ forms a triadic closure, which indicates strong social relations~\cite{simmel1908sociology}. This example shows that it is a problem for existing HIN-based RS methods to assume that nodes of the same type have the same weight when computing meta-path based similarities.

\begin{figure}
	\centering
	\includegraphics[width=0.7\columnwidth]{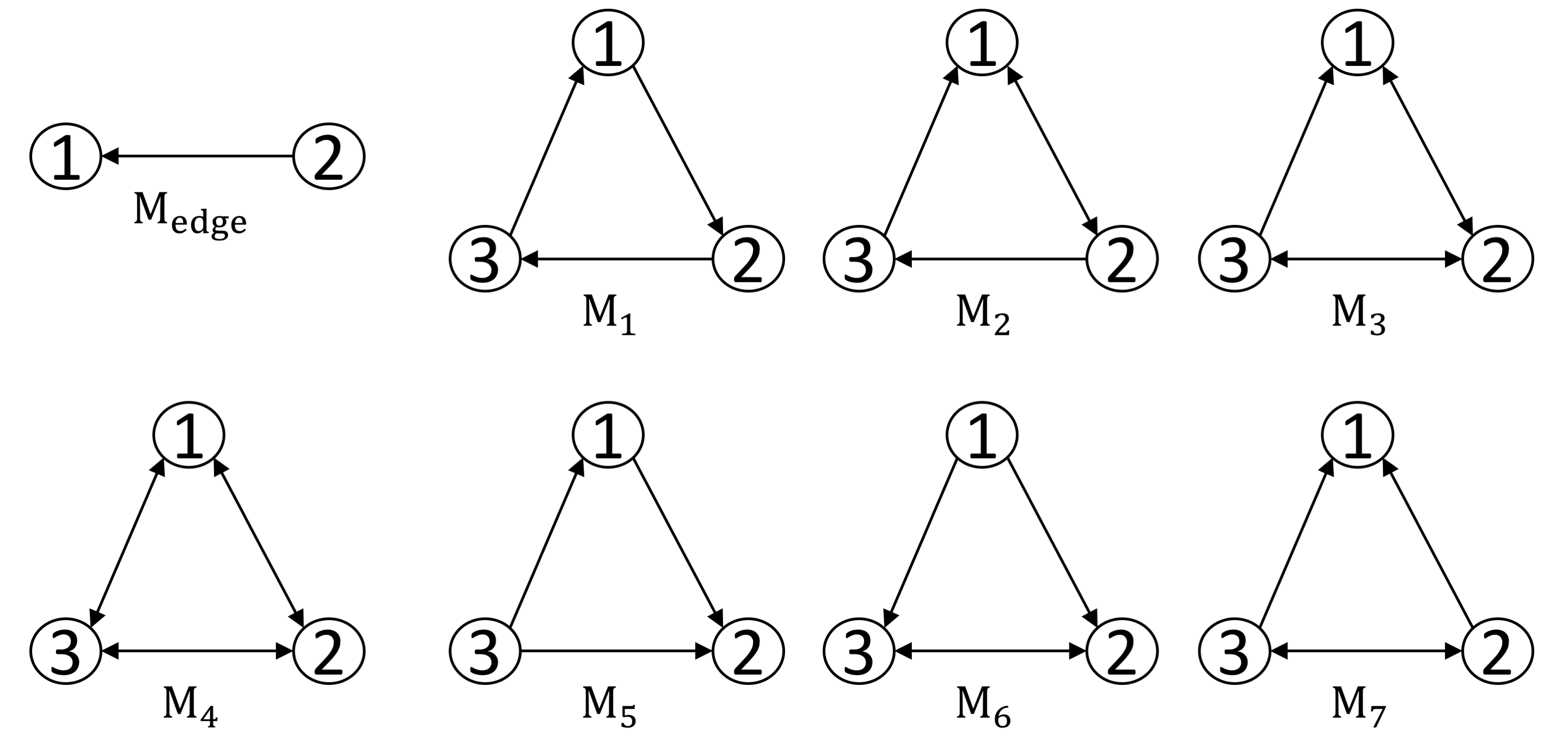}
	\caption{Typical 3-node motifs. Note that 1, 2, and 3 represent the positions of a motif where a node can occur.}
	\label{fig-motif-example}
	\vspace{-0.2in}
\end{figure}

In the literature, the triangle, formed by $u_1, u_2,$ and $u_3$, can be generalized as \textit{network motif}, which is a local structure involving multiple nodes in a homogeneous graph, e.g. social graph.  For example, in Figure~\ref{fig-motif-example}, we show seven typical 3-node motifs. Proposed in~\cite{milo2002network}, motif has been demonstrated to be a very important local structure underlying various complex networks. It is also called \textit{higher-order} relations in the literature~\cite{benson2016higher}. We call the connection directly connecting nodes of same type edge-based \textit{first-order} relations. In~\cite{benson2016higher,zhao2018ranking}, motif has been shown to be very important in obtaining the similarities among nodes of a graph. However, no previous works have explored the influence of motif in HIN. In this paper, we firstly propose Motif Enhanced Meta-Path (MEMP) to incorporate motif-based higher-order relations into conventional meta-path based similarity computation, then design the motif HIN-based RS (MoHINRec) for the recommendation. We then conduct experiments on two real-world datasets, Epinions and CiaoDVD, to demonstrate the effectiveness of motif on HIN-based RSs. The code of the proposed MoHINRec is available in https://github.com/HKUST-KnowComp/MoHINRec.

\section{Framework}
\label{sec-framework}

In this section, we present the details of our proposed framework for integrating Motif into HIN-based Recommender (MoHINRec). 

\subsection{Meta-path based similarity computation}

In previous HIN-based scenarios, meta-paths are used to capture the complex semantics underlying the similarities between nodes of any types. In this part, we give a brief introduction on the counting-based meta-path based similarity. Given a meta-path, we want to compute the similarities between the source and the target nodes, i.e., users and items (Business) in Figure~\ref{fig-example-hin}(b). 
Commuting matrix~\cite{sun2011pathsim} has been used to compute the counting-based similarity matrix of a meta-path.
Suppose we have a meta-path $\mathcal{P} = (A_1,A_2,\ldots,A_l)$, where $A_i$'s are node types in $\mathcal{A}$, which represents the entity type set $\cA$ in a HIN.
We can define a matrix $\bW_{A_iA_j}$ as the adjacency matrix between node type $A_i$ and node type $A_j$.
Then the commuting matrix for meta-path $\mathcal{P}$ is $\bC_{\cP} =\bW_{A_1,A_2}\cdot \bW_{A_2,A_3}\cdot ...\cdot \bW_{A_{l-1},A_l}$, which represents the number of instances of $\cP$ connecting two nodes of type $A_1$ and $A_l$.
For example, in Figure~\ref{fig-example-hin}(b), the commuting matrix for $\cP_2$ can be obtained by $\bC_{\cP_2} = \bW_{UU} \cdot \bW_{UB}$, where $\bW_{UB}$ is the adjacency matrix between type $U$ and type $B$, and $\bW_{UU}$ is the adjacency matrix for  type $U$, i.e., recording the relations among all instance nodes of type $U$.
This shows that counting-based similarities for a meta-path can be computed by multiplying a sequence of adjacency matrices. In this paper, we adopt the counting-based similarity for users and items given a meta-path. In practice, we can implement this procedure in a very efficient way if the adjacency matrices $\bW$'s are sparse.

\subsection{MoHINRec Framework}
In this part, we elaborate on the MoHINRec framework, consists of motif-based adjacency matrix, MEMP based similarity computation, and recommendation model with FMG~\cite{zhao2017meta}.


Firstly, we give the definition of motif-based adjacency matrix in the context of HIN.~\footnote{For the formal definition of motif, we refer readers to~\cite{bensonsupplementary}.} Given a motif $\mM_k$, the definition of  motif-based adjacency matrix for node type $A_i$ in a HIN is:
\begin{equation}
(\bW^{A_i}_{M_k})_{ij} = \sum\limits_{v_i,v_j \in \cV, \phi(v_i) = \phi(v_j) = A_i} \b1(v_i, v_j \text{ occur in } M_k),
\end{equation}
where $i \neq j$, $v_i$ and $v_j$ belong to type $A_i$, and $\b1(s)$ is the truth-value indicator function, i.e., $\b1(s) = 1$ if the statement $s$ is true and 0 otherwise. Note that the weight is added to $(\bW^{A_i}_{M_k})_{ij}$ only if node $v_i$ and $v_j$ occur in the given motif $M_k$.  Note that for the 3-node motifs in Figure~\ref{fig-motif-example}, there are six cases where two nodes occur in a 3-node motif because the graphs are directed. Therefore, computing the motif-based adjacency matrix incurs subgraph counting. Fortunately, for the seven motifs, there are simple formulas to compute the corresponding motif-based adjacency matrices. Here, we omit the detail for clarity, and refer the readers to~\cite{bensonsupplementary,zhao2018ranking} for the formulas.

\begin{figure}
	\centering
	\subfigure[Example graph for type $A_i$.]{\includegraphics[width=0.2\textwidth]{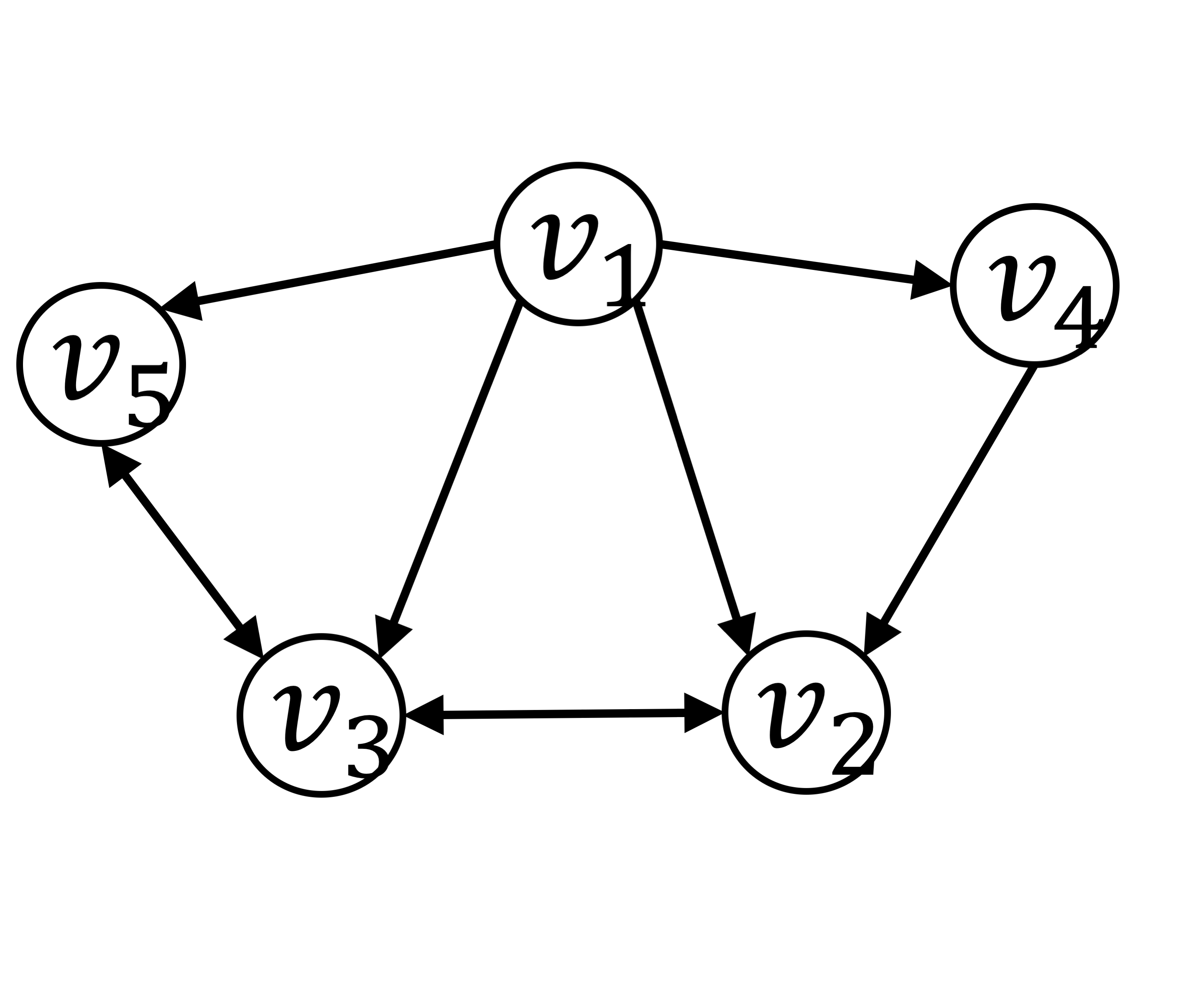}}
	\subfigure[$\bW^{A_i}_{M_6}$.]{\includegraphics[width=0.2\textwidth]{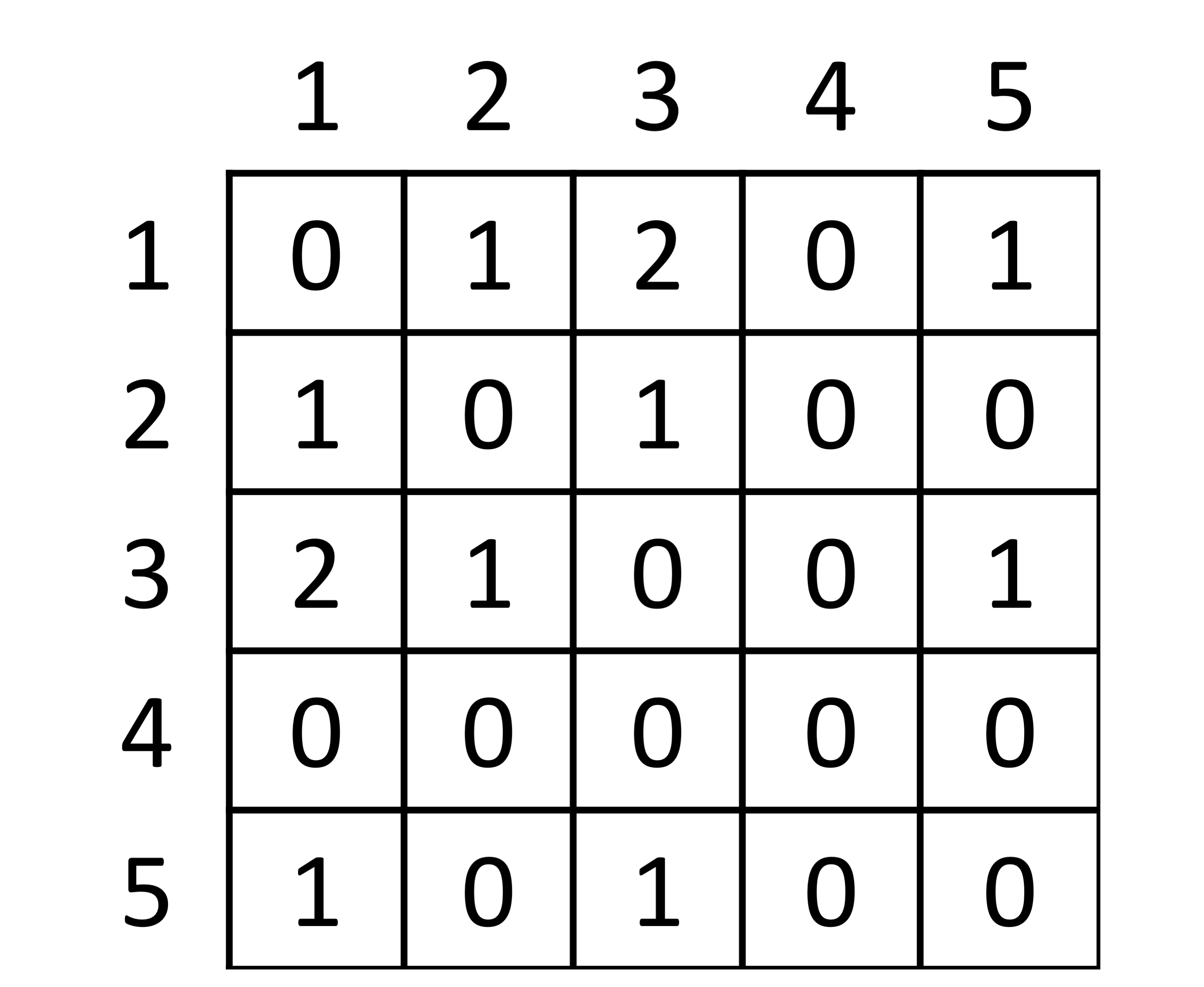}}
	\caption{(a) is an example graph generated by the adjacency matrix $\bW_{A_iA_i}$. (b) is the motif-based adjacency matrix for motif $M_6$ in Figure~\ref{fig-motif-example}.}
	\label{fig-example-motif-adj}
	\vspace{-0.2in}
\end{figure}

Now, we give an example to illustrate the motif-based adjacency matrix in Figure~\ref{fig-example-motif-adj}. In Figure~\ref{fig-example-motif-adj}(a), 5 nodes $v_1, v_2, ..., v_5$ belong to type $A_i$, and the relations among them are recorded by $\bW_{A_iA_i}$. Thus, an example graph of node type $A_i$ can be drawn accordingly in Figure~\ref{fig-example-motif-adj}(a). Assume that we want to compute the motif-based adjacency matrix given the motif $M_6$ in Figure~\ref{fig-motif-example}. The result $\bW^{A_i}_{M_6}$ is shown in Figure~\ref{fig-example-motif-adj}(b). Details of the computation are given in~\cite{bensonsupplementary,zhao2018ranking}. From the matrix, we can see that $(\bW^{A_i}_{M_6})_{13} = (\bW^{A_i}_{M_6})_{31} = 2$, because $v_1$ and $v_3$ occur in two instances of $M_6$, i.e., the triangles formed by $(v_1, v_3, v_5)$ and $(v_1, v_2, v_3)$. This example explains the meaning of motif-based adjacency matrix, i.e., it records the frequency of two nodes occurring in a given motif.

Given a meta-path $\cP$, for which the commuting matrix can be obtained by $\bC_P =\bW_{A_1,A_2}\cdot \bW_{A_2,A_3}\cdot ...\cdot \bW_{A_{l-1},A_l}$. Assuming $A_i$ and $A_{i+1}$ are of the same type, we can construct the motif-based adjacency matrix for node type $A_i$, denoted as $\bW^{A_i}_{M_k}$, given a motif $M_k$. Then,  same as~\cite{zhao2018ranking}, we propose to use linear combination to fuse the edge-based and motif-based adjacency matrices for node type $A_i$. Specifically, we generate the MEMP-based adjacency matrix as follows.
\begin{equation}
\label{eq-combination}
\bH^{A_i}_{M_k} = (1 - \alpha) \cdot \bW_{A_iA_{i+1}} +  \alpha \cdot \bW^{A_i}_{M_k},
\end{equation}
where $\alpha \in [0, 1]$ balances the combination of edge-base and motif-based adjacency matrices. When $\alpha = 0$, it means we only use the type adjacency matrix, i.e., the edge-based adjacency matrix for node type $A_i$, and when $\alpha = 1$, it means we use the motif-based adjacency matrix alone.
Then, in the computation of commuting matrix $\bC_P$, we replace $\bW_{A_iA_{i+1}}$ with $\bH^{A_i}_{M_k}$. In this way, we incorporate motif into meta-path based similarity computation. For each meta-path, $\cP_l$, we can obtain a new commuting matrix $\hat{\bC}_{\cP}$, recording the number of the MEMP instance connecting users and items. In this work, we use the frequency, i.e., $(\hat{\bC}_{P_l})_{ij}$, to denote the similarity between user $u_i$ and item $b_j$ under the MEMP $\cP_l$. 

%

After we obtain $L$ MEMP-based similarity matrices, we adopt a state-of-the-art HIN-based RS method~\cite{zhao2017meta}, which first factorizes each similarity matrix separately to obtains a group of user and item latent features from each matrix, 
and then feeds the features to a factorization machine (FM)~\cite{rendle:tist2012} to generate recommendations.

\begin{table}[ht]
	\centering
	\caption{Statistics of Datasets.}
	\vspace{-0.1in}
	\label{tb-dataset-stat}
	\begin{tabular}[\columnwidth]{c|cccc}
		\toprule
		& Users & Items & Ratings &  Social relations  \\ \midrule
		Epinions &   22,164 &  296,277 &   912,441 & 355,754  \\ \hline
		CiaoDVD &  17,615 & 16,121  & 72,345  &40,132  \\ \bottomrule
	\end{tabular}
\end{table}

%

\section{Experiment and Analysis}
In this section, we present the experimental results.


\subsection{Experimental Settings}
\textbf{Datasets.} We conduct experiments on two real-world datasets: Epinions and CiaoDVD, which are review websites where users can write reviews on products and rate the reviews of other users. Moreover, users can add other users as trustworthy users if they like their reviews. The Epinions dataset is provided by~\cite{tang-etal12b}, and CiaoDVD is provided by~\cite{guo2014etaf}. The statistics of the two datasets are shown in Table~\ref{tb-dataset-stat}. Note that although the datasets include other SIs such as the categories and reviews of items, the only relation that exists between nodes of same type is social relation. Thus, in our experiments, we use meta-paths $\cP_1$ and $\cP_2$ in Figure~\ref{fig-example-hin}(b) to demonstrate the effectiveness of MEMP and omit the statistics of the other SIs in Table~\ref{tb-dataset-stat}.


\noindent\textbf{Evaluation Metrics.} We choose to tackle the rating prediction task, which is widely used to evaluate CF-based RSs. We choose two evaluation metrics, Mean Absolute Error (MAE) and Root Mean Square Error (RMSE), to evaluate our framework. 


\noindent\textbf{Baselines.} We compare our proposed models with the following RS methods:
\begin{itemize}
	\item \textbf{RegSVD}~\cite{paterek2007improving}: It is the standard matrix
	factorization method with $\ell_2$ regularization. We use the implementation in an open source library LibRec~\cite{guo2015librec}.
	\item \textbf{SoReg}~\cite{ma2011recommender}: It is a MF-based method employing social connections as regularization terms. We use the implementation in an open source library LibRec~\cite{guo2015librec}.
	\item \textbf{SocialMF}~\cite{jamali2010matrix}: It is a MF based framework which incorporates social trust propagation. We use an open source library LibRec~\cite{guo2015librec} in the implementation.
	\item \textbf{FMG}~\cite{zhao2017meta}: It is a state-of-the-art HIN-based RS method adopting the ``MF+FM'' framework. We use the implementation given by the authors. For this method, we choose the meta-paths $\cP_1$ and $\cP_2$ in Figure~\ref{fig-example-hin}(b) without incorporating motifs.
\end{itemize}

Note that there are other HIN-based RS methods for rating prediction~\cite{yu2014personalized,shi2015semantic}, but FMG has been shown to be consistently superior~\cite{zhao2017meta} to these methods. Therefore, FMG is used in the experiments as the state-of-the-art baseline representing HIN-based RSs. Our MoHINRec method is based on FMG with different MEMP-based similarities. For a given motif $M_k$ in Figure~\ref{fig-motif-example}, we denote the method as MoHINRec($M_k$). In this paper, we mainly focus on 3-node motifs, i.e., the seven motifs shown in Figure~\ref{fig-motif-example}.

\textbf{Settings.} For the experimental settings, we randomly split each dataset into training, validation and test data with a ratio of 8:1:1. In the training process, the training data is used to fit the model, and the validation data is used to choose the best parameters, and the test data is used to compute the prediction errors of the models. For RegSVD, SoReg, and SocialMF, the rank is set to be $10$, and for FMG, we adopt the same settings as~\cite{zhao2017meta}, i.e., $F=10, K=10$. The regularization parameters $\lambda_w, \lambda_v$ and combination factor $\alpha$ in Eq.~\eqref{eq-combination} are tuned by the validation data. Note that for simplicity, we set $\lambda_w = \lambda_v = \lambda$. We repeat each experiment five times by randomly splitting the datasets and report the average results.

\begin{table}[]
	\centering
	\caption{Performance comparison of different methods. The best performance in each column is highlighted in bold.}
		\vspace{-0.1in}
	\label{tb-performance-comparison}
	\begin{tabular}{c|c|c|c|c}
		\toprule
		& \multicolumn{2}{c|}{Epinions} & \multicolumn{2}{c}{CiaoDVD} \\ \cline{2-5}
		& RMSE & MAE & RMSE & MAE \\ \midrule
		RegSVD & 1.9665 & 1.50845 & 2.2396 & 1.8017 \\ 
		SocialMF & 1.1349 & 0.8917 & 1.0203 & 0.7987 \\ 
		SoReg & 1.6783 & 1.4033 & 1.1973 & 0.9363 \\ 
		FMG & 1.0857 & 0.8366 & 1.0108 & 0.7836 \\ \midrule
		MoHINRec($M_1$) & 1.0807 & 0.8332 & 1.0000 & 0.7730 \\ 
		MoHINRec($M_2$) & 1.0763 & 0.8316 & 0.9935 & 0.7677 \\ 
		MoHINRec($M_3$) & \textbf{1.0740} & \textbf{0.8291} & 0.9869 & \textbf{0.7659} \\ 
		MoHINRec($M_4$) & 1.0768 & 0.8318 & 0.9871 & 0.7723 \\
		MoHINRec($M_5$) & 1.0759 & 0.8306 & 0.9869 & 0.7738 \\
		MoHINRec($M_6$) & 1.0806 & 0.8346 & 0.9868 & 0.7736 \\ 
		MoHINRec($M_7$) & 1.0776 & 0.8318 & \textbf{0.9867} & 0.7690 \\ \bottomrule
	\end{tabular}
\end{table}

\subsection{Performance Comparison}
\label{subsec-rmse-comp}
We show the performance of all methods in terms of RMSE and MAE in Table~\ref{tb-performance-comparison}. We can see that MoHINRec with any of the seven motifs outperforms all baselines consistently on both datasets. This demonstrates the effectiveness of MEMP for HIN-based recommendation. Besides, the best performance is achieved when $\alpha$ is in $(0,1)$. For example, on Epinions, the lowest RMSEs are obtained when $\alpha =0.6$ for $M_6$, $\alpha = 0.1$ for $M_3$ and $\alpha=0.5$ for $M_5$. On CiaoDVD, the lowest RMSEs are obtained when $\alpha =0.2$ for $M_5$, $\alpha = 0.5$ for $M_6$ and $\alpha=0.5$ for $M_7$. This aligns with our assumption that motif-based higher-order relations and edge-based first-order relations among nodes of the same type are complementary to each other in meta-path based similarity computation. The same observation has been reported in~\cite{zhao2018ranking} for the user-ranking task in social networks. We point out below two observations from Table~\ref{tb-performance-comparison}.

First, the performance gain of MoHINRec varies across different datasets and different motifs. On Epinions, the best RMSE and MAE are achieved by $M_3$, while on CiaoDVD, the best performance is obtained by $M_7$ for RMSE and $M_3$ for MAE. The performance gains of the other methods vary. It means that despite the usefulness of MEMP for HIN-based RS methods, the performance gain is dependent on the motif and dataset. 

Second, FMG, the state-of-the-art HIN-based RS method, outperforms all other MF-based methods, i.e., RegSVD, SoReg, and SocialMF. It demonstrates the power of the ``MF+FM'' framework. However, MoHINRec clearly beats FMG. On Epinions, MoHINRec with $M_3$ decreases RMSE from $1.0857$ to $1.0740$, and on CiaoDVD, MoHINRec with $M_7$ decreases RMSE from $1.0108$ to $0.9867$. It means by incorporating motif into exiting meta-path based similarity computation, we can further improve the recommending performance. 

From the experimental results, we can see that motif-based relations can benefit meta-path based recommendation with FMG model.

\section{Related Work}
\label{sec-related}
In this section, we review related work on HIN-based RSs and motif in homogeneous graphs.

\subsection{Recommendation in HIN}

To better make use of rich side information in RSs, HIN has been proposed to represent disparate heterogeneous information into a single graph. Based on meta-path, several approaches have been proposed to exploit HIN for the recommendation task.~\cite{yu2014personalized,shi2015semantic,zhao2017meta,shi2018heterogeneous,han2018aspect,Hu2018LMB,Wang2018Billion,zhao2018learning,Fan2019MHG} However, all the HIN-based RS methods ignore the motif-based higher-order relations among nodes of same type.


\subsection{Motif in homogeneous graphs}
Motif can be used to characterize higher-order relations in homogeneous graphs.
Network motif was first introduced in~\cite{milo2002network}.
It has been shown to be useful in many applications such as social networks~\cite{rotabi2017detecting}, and temporal networks~\cite{paranjape2017motifs}. 
Recently, it was shown that motif can also be used for graph clustering or community detection~\cite{benson2016higher,yin2017local}, user analysis~\cite{zhang2017structinf} and ranking~\cite{zhao2018ranking} in social networks. 
Compared to these previous studies, which are all in homogeneous graphs, we are the first to incorporate the motif-based higher-order relations into heterogeneous graphs.

\section{Conclusion and Future work}
\label{sec-conclusion}
In this paper, we explore motif-based higher-order relations in HIN-based RSs, which are proved to be useful in  homogeneous graphs of various domains. 
We propose the motif-enhanced meta-path (MEMP) for computing the similarities between users and items in HIN, and experimental results on two real-world datasets, Epinions and CiaoDVD, demonstrate that the proposed MoHINRec built on MEMP-based similarities is superior to existing HIN-based RS methods. For future work, we will explore motif-based relations among nodes of different types. This may lead to novel structures that can generalize meta-path and meta-graph in HIN.
\vspace{-0.1in}


\section{Acknowledgments}
Dik Lun Lee and Huan Zhao are supported by the Research Grants Council HKSAR GRF (No. 16215019).
Yangqiu Song and Yingqi Zhou are supported by the Early Career Scheme (ECS, No. 26206717) from Research Grants Council in Hong Kong. We also thank the anonymous reviewers for their valuable comments and suggestions that help improve the quality of this manuscript.
\vspace{-0.1in}
\bibliographystyle{ACM-Reference-Format}
\bibliography{draft}

\end{document}